\documentclass{article}
\usepackage{ijcai22}

\usepackage{times}
\usepackage{soul}
\usepackage{url}
\usepackage[hidelinks]{hyperref}
\usepackage[utf8]{inputenc}
\usepackage[small]{caption}
\usepackage{graphicx}
\usepackage{amsmath}
\usepackage{amsthm}
\usepackage{booktabs}
\usepackage{algorithm}
\usepackage{algorithmic}
\usepackage{stfloats}
\usepackage{amssymb}
\usepackage{textcomp}
\usepackage{xcolor}
\usepackage{subfigure}
\usepackage{amsfonts}
\usepackage{multirow}

\title{Hyperbolic Self-supervised Contrastive Learning Based Network Anomaly Detection}

\author{
    Yuanjun Shi
    \affiliations
    Tianjin University
    \emails
    switchsyj@tju.edu.cn
}

\begin{document}
\maketitle
\begin{abstract}
Anomaly detection on the attributed network has recently received increasing attention in many research fields, such as cybernetic anomaly detection and financial fraud detection. With the wide application of deep learning on graph representations, existing approaches choose to apply euclidean graph encoders as their backbone, which may lose important hierarchical information, especially in complex networks. To tackle this problem, we propose an efficient anomaly detection framework using hyperbolic self-supervised contrastive learning. Specifically, we first conduct the data augmentation by performing subgraph sampling. Then we utilize the hierarchical information in hyperbolic space through {exponential mapping} and {logarithmic mapping} and obtain the anomaly score by subtracting scores of the positive pairs from the negative pairs via a discriminating process. Finally, extensive experiments on four real-world datasets demonstrate that our approach performs superior over representative baseline approaches.

\end{abstract}

\vspace{-4mm}
\section{Introduction}

\begin{figure}[t]
    \centering
    \subfigure[Attribute Anomaly]{
        \label{fig:attr}
        \includegraphics[scale=0.09]{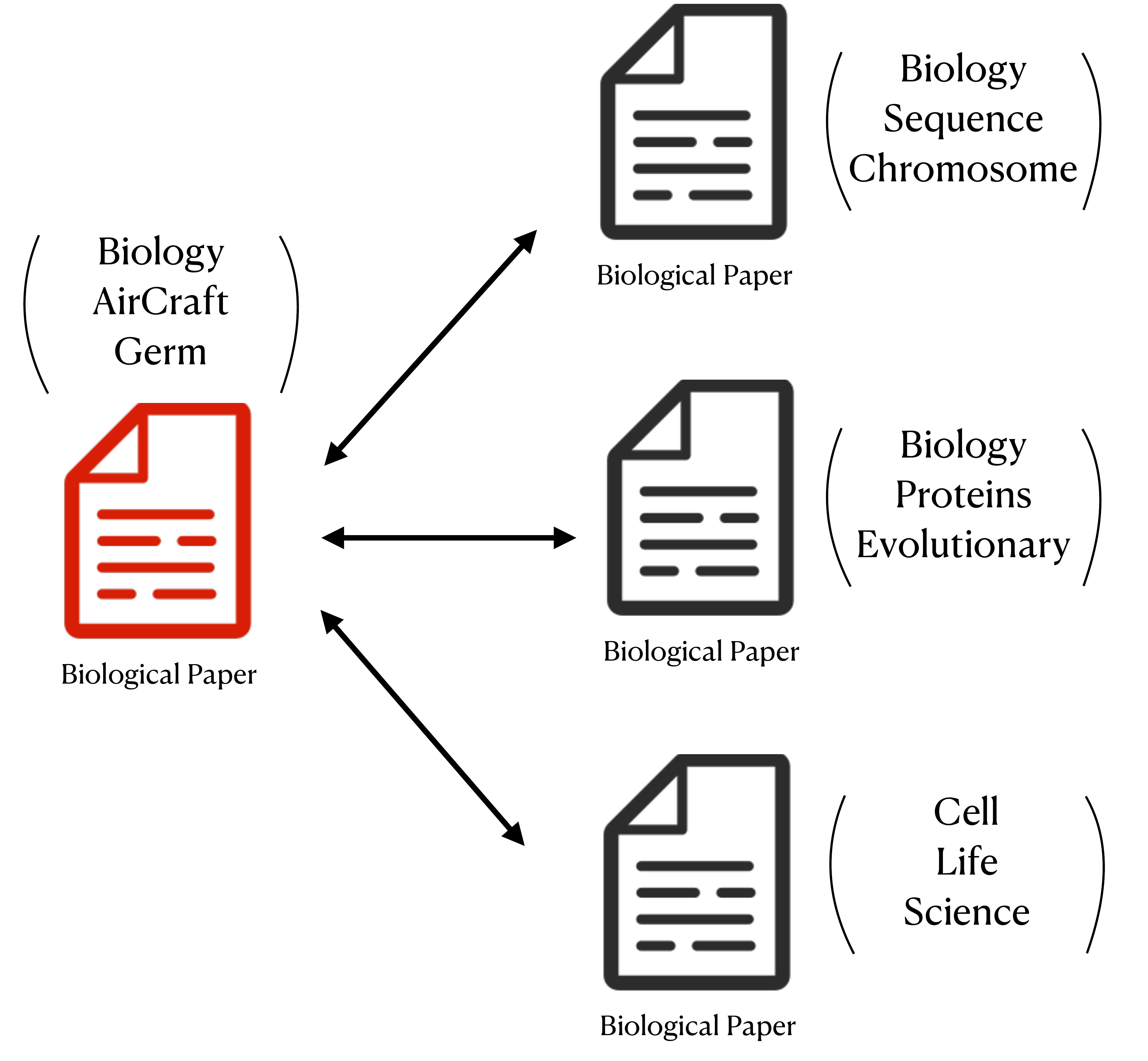}
    }
    \subfigure[Structural Anomaly]{
        \label{fig:struct}
        \includegraphics[scale=0.09]{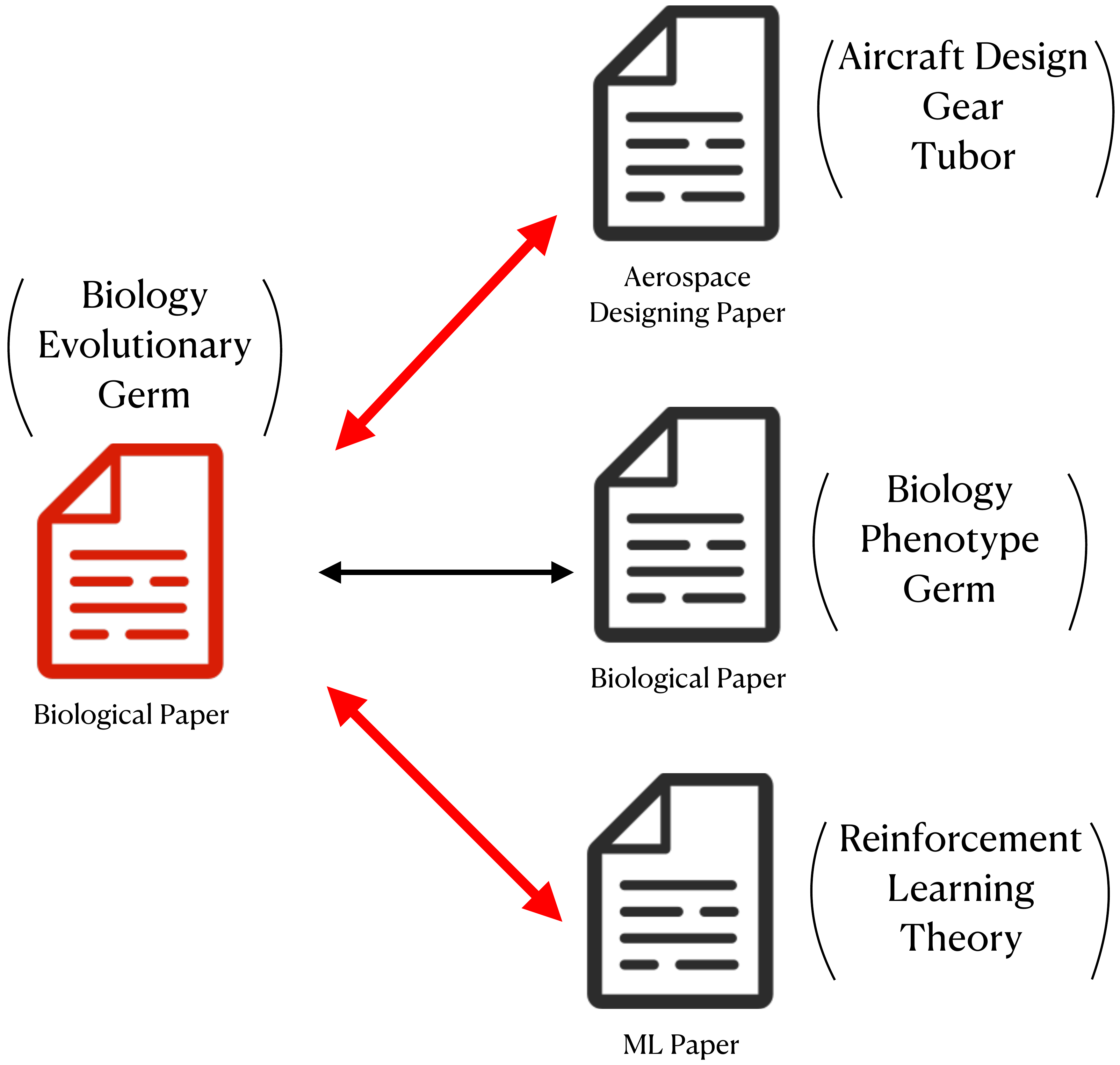}
    }
    \vspace{-2mm}
    \caption{(a) For the attribute anomaly in a citation network, the red anomaly paper is perturbed with its word attributes.  (b) For the structural anomaly, the anomalous citation relationships are indicated by the red edges.}
    \label{fig}
    \vspace{-5mm}
\end{figure}

Anomaly detection on attributed networks has received enormous attention. Anomaly detection~\cite{li2019specae,ding2019dominant} refers to the process of spotting data instances that deviate significantly from the majorities. Since the attributed graph is made up of attribute features and edges, in general, there are two types of anomalies, i.e. the attributed anomaly shown in Figure \ref{fig:attr} and the structural anomaly shown in Figure \ref{fig:struct}. As a result, due to its complexity, it seems difficult to apply a simple anomaly detection method, which is initially employed for attribute-only data~\cite{chen2001ocsvm} or static plain networks~\cite{duan2020aane}. Furthermore, accessing the ground-truth label in a corrupted attributed graph is expensive. Hence, it is difficult to identify abnormalities in an attributed network immediately due to a lack of appropriate supervision signals. Hence, the key to addressing this issue is to conclude the patterns of normal nodes or abnormal nodes in diverse patterns, and utilize the information from the corrupted network without much supervision signals.

Several approaches have been made to deal with anomaly detection for attributed networks. Some are shallow learning methods, including ~\cite{perozzi2016amen,li2017radar,peng2018anomalous,xu2007scan}. Unfortunately, shallow models are limited by their ability to capture specific patterns due to data sparsity or complex modality issues~\cite{ding2019dominant}. For example, ~\cite{xu2007scan} mainly focuses on graph node attributes or topological structures. ~\cite{perozzi2016amen} takes both into account, but only considers its neighbors rather than the node itself.

Moreover, with the development of graph deep learning techniques, graph neural network(GNN) becomes a powerful tool for graph modeling, exhibiting superior performance in a variety applications. For example, ~\cite{kipf2017gcn} proposes graph convolution network by applying the convolutional operation on graph data. ~\cite{velickovic2018gat} employs the attention mechanism on graph, which is more competitive in specific inductive tasks. Specifically, some approaches for anomaly detection try to learn feature representation of normality using graph encoders in a generative paradigm, including~\cite{li2019specae,ding2019dominant,fan2020anomalydae}. The assumption for this is that normal instances are more likely to be better reconstructed than abnormal ones ~\cite{pang2021preview}. However, first, this reconstruction process aims to data compressing rather than anomaly detection, which is designed to learn the general representation of the underlying normal instances. Second, it takes great cost in the computation of reconstruction process when the network is large, as graph convolution operation for the whole graph is required (e.g. SpecAE~\cite{li2019specae}).

Self-supervised contrastive learning is a feasible alternative for overcoming the aforementioned limitations. Contrastive learning has shown its great potential in many application fields, including computer vision~\cite{chen2020simclr,he2020moco}, natural language processing~\cite{gao2021simcse} and graph representation learning~\cite{Zhu:2020grace,zhu2021gca,mvgrl}. It is intended to optimize contrastive loss in order to learn the generic representation of different types of instances. Specifically, similar positive pairs are pulled together, while dissimilar negative ones are pushed apart. 

Inspired by \cite{liu2021cola}, there is consistency between a normal node and its neighbors. While for an abnormal node, there is inconsistency with neighbors in its attributes, graph topologies, or both. Furthermore, the prediction score of the contrastive discriminator is utilized to calculate contrastive loss, which is highly related to alignment and uniformity between different pairs. In this view, contrastive learning is inherently helpful to solve anomaly detection tasks by aligning nodes with their neighbors, where abnormal nodes fail to follow this pattern. Despite their effectiveness, recent work~\cite{wilson2014spherical} has demonstrated that graph data shows more hyperbolic geometric properties, as the number of nodes rises exponentially with the growth of distance to the root of the whole graph. Intuitively, complex networks with hierarchical information can be naturally embedded into hyperbolic space. Recent approaches ~\cite{hgnn,chami2019hyperbolicgcn,zhang2021hyperbolicgat} have exhibited great modeling ability of hyperbolic graph neural networks on attributed graphs, resulting in great improvements on several node classification tasks.  

In view of this, we extend the contrastive learning method to the hyperbolic space via the self-supervised learning paradigm to utilize the hierarchical relationship between different nodes and to solve accompanying challenges: 
\textit{(1) how to sample positive and negative pairs, which are subsequently encoded into the hyperboloid model making use of hierarchical information; (2) how to distinguish between positive and negative pairs in the hyperboloid model, which is different from the fundamental contrastive learning framework.} The main contributions are summarized as below. 

\begin{itemize}
    \item Propose an efficient anomaly detection framework using hyperbolic self-supervised contrastive learning. To the best of our knowledge, we are the first to extend contrastive learning to hyperbolic space for anomaly detection utilizing the rich information of hierarchical graph data via \textit{exponential mapping} and \textit{logarithmic mapping}.
    
    \item 
    Perform subgraph sampling and target nodes anonymization to conduct data augmentation, and employ discriminator after hyperboloid encoding and tangent space decoding. To demonstrate the notion of the hierarchical information, we further analyze the key role of our hyperboloid encoder and tangent decoder in {the ablation study}. Furthermore, we also conduct subgraph matrix calculation stably and efficiently.
    \item Conduct extensive experiments on four real-world datasets and demonstrate that our approach performs superior over representative baseline approaches. 
    Moreover, results suggest that the more hyperbolic the dataset is, the more improvements this method can make.
\end{itemize}



\section{Related Work}

\subsubsection{Anomaly Detection on Networks}

Traditional non-deep learning anomaly detection techniques, such as AMEN~\cite{perozzi2016amen} focus on detecting abnormality of subgraphs with node information in an ego-net. Radar~\cite{li2017radar} learns residual error by reconstruction and considers the top-k nodes with larger norms of error as anomalies. 
ANOMALOUS~\cite{peng2018anomalous} tries to use CUR matrix decomposition to select attributes, which is closely related to network structure, followed by the same residual analysis used in Radar. Although these techniques try to capture graph information from attributes and structures, they cannot work well on large complex networks.

Moreover, graph deep learning techniques have been applied in varieties of applications. DOMINANT~\cite{ding2019dominant} aims to minimize network and attribute reconstruction errors with node representations embedded by a graph convolutional encoder. 
SpecAE~\cite{li2019specae} applies graph convolutional encoder to learn nodes representations and detect anomalies through a density estimation approach. 
Unfortunately, reconstruction-based methods also have disadvantages:  the training object is reconstruction error instead of anomaly score itself, which results in overfitting to the general patterns of normal nodes. Hence, it cannot handle the out-of-distribution(OOD) problem in inductive tasks. 

More recent works combine supervised contrastive learning with anomaly detection. 
Contrastive learning across different data augmentation views has recently achieved the state-of-the-art in the field of self-supervised learning, aiming to learn representations by pulling similar pairs closer and pushing dissimilar ones away.  
Such as CoLA~\cite{liu2021cola} uses the ``target node v.s local subgraph" view. 
ANEMONE~\cite{jin2021anemone} proposes a multi-view contrastive learning method, which employs not only ``node v.s neighbor" but also ``node v.s node".  Despite their success, it is not an appropriate way to model large complex networks, which contains rich hierarchical information. 

\vspace{-1mm}
\subsubsection{Hyperbolic Graph Neural Network}
Recently, Hyperbolic Graph Neural Networks (HGNNs~\cite{hgnn}) has attracted more attention. Different from GCN~\cite{kipf2017gcn} and GAT~\cite{velickovic2018gat}, HGNNs are designed for modeling complex networks with higher ordered hierarchical information~\cite{wilson2014spherical}. 
HGCN~\cite{chami2019hyperbolicgcn} defines convolution operation and neighborhood aggregation in hyperbolic space, whereas HAT~\cite{zhang2021hyperbolicgat} proposed a graph attention mechanism in hyperbolic space. And these hyperbolic models have shown great improvements on graph modeling tasks.
To tackle the problem mentioned above, in this paper, we apply contrastive learning method for anomaly detection in the hyperbolic space while paying extra attention to the hierarchical information between nodes.

\section{Proposed Method}

\begin{figure*}[t]
    \centering
    \includegraphics[scale=0.43]{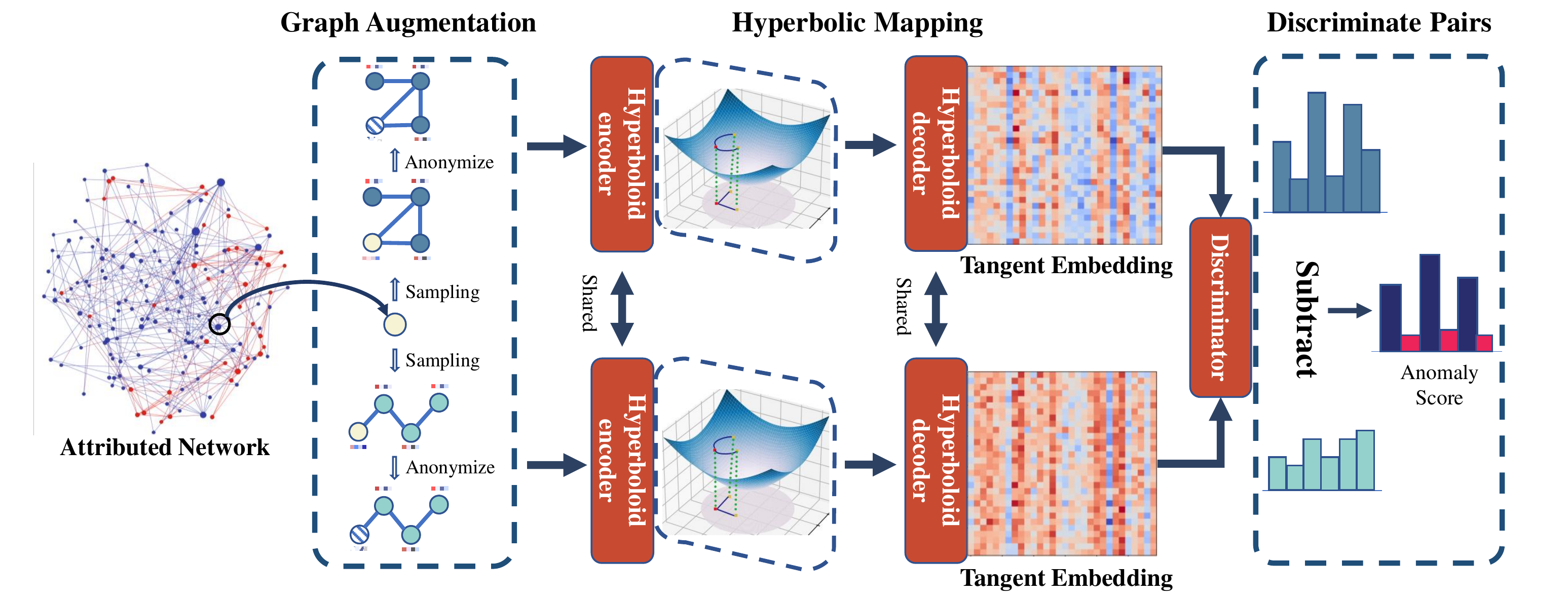}
    \vspace{-3mm}
    \caption{Our model framework: we first perform subgraph sampling and target nodes anonymization from the real-world attributed network containing anomalies as \textbf{Graph Data Augmentation} step, and then apply \textbf{Hyperbolic Mapping} through \textit{exponential mapping} and \textit{logarithmic mapping}. Finally, we can get anomaly score by subtracting scores of the positive pairs from the negative pairs via the \textbf{Discriminating Process}. Specifically, we should note that red nodes denote the anomalies while blue nodes denote the normal ones in the attributed network. }
    \label{fig:model}
    \vspace{-3mm}
\end{figure*}

\subsection{Preliminaries: Hyperbolic Geometry}
\paragraph{Riemannian Manifold.} A manifold $\mathcal{M}$ is a topological space mapping from the original space $\mathbb{R}^d$ near each point, referring to a generalization of surfaces in high dimensions. Intuitively, for each $x\in\mathcal{M}$, there is a tangent space $\mathcal{T}_\mathbf{x}\mathcal{M}$ of the same dimension as $\mathcal{M}$, including all possible directions of curves in $\mathcal{M}$ passing through $\mathbf{x}$.

For a smooth manifold $\mathcal{M}$, a Riemannian manifold is defined as a pair ($\mathcal{M}, \textbf{g}$), where $\textbf{g}=(g_\mathbf{x})_{\mathbf{x}\in\mathcal{M}}$ is a Riemannian metric measuring the distance on $\mathcal{M}$, and $g_\mathbf{x}:\mathcal{T}_\mathbf{x}\mathcal{M}\times\mathcal{T}_\mathbf{x}\mathcal{M}\to\mathbb{R}$. We can use exponential and logarithmic methods to map between Riemannian manifold and its tangent space. For a certain point $\mathbf{x}\in\mathcal{T}_\mathbf{x}\mathcal{M}$, the \textit{exponential mapping}: $exp_\mathbf{x}(\mathbf{u})$ projects the vector from tangent space to Riemannian manifold. While the \textit{logarithmic mapping}: $log_\mathbf{x}(\mathbf{v})$ projects the vector to its tangent space. For two given points $\mathbf{x}, \mathbf{y}\in\mathcal{M}$, the \textit{parallel transport} $P_{\mathbf{x}\to \mathbf{y}}(\mathbf{u})$ maps a vector $\mathbf{u}\in\mathcal{T}_\mathbf{x}\mathcal{M}$ to $\mathcal{T}_\mathbf{y}\mathcal{M}$.
\paragraph{Hyperbolic Space.} The hyperbolic space is a connected Riemannian manifold with negative curvature. There are several hyperbolic models, such as Poincaré or Hyperboloid model (a.k.a the Minkowski or Lorentz model). We will give the detailed information about hyperboloid model utilized for further discussion.

Given two vectors $\mathbf{u}, \mathbf{v}$ in the tangent space  $\mathcal{T}_\mathbf{x}\mathbb{H}_d^K=\{\mathbf{y}\in\mathbb{R}^{d+1}|\langle\mathbf{x}, \mathbf{y}\rangle_\mathcal{L}=0\}$, we denote the hyperboloid metric $\mathbf{g}_\mathbf{x}^K(\mathbf{u}, \mathbf{v})=\langle\mathbf{u},\mathbf{v}\rangle_\mathcal{L}$, 
where $\langle{\cdot, \cdot}\rangle_L: \mathbb{R}^{d+1}\times\mathbb{R}^{d+1}\to\mathbb{R}$ denotes the Minkowski inner product:
\begin{equation}
    \langle\mathbf{x}, \mathbf{y}\rangle_{\mathcal{L}}=-x_0y_0+x_1y_1+...+x_dy_d.
\end{equation}
Hence, the hyperboloid model $\mathbb{H}$ with a negative curvature $-1/K(K>0)$ in $d$ dimensions is defined as:
\begin{equation}
    (\mathbb{H}^{d,K}, \textbf{g})= \{\mathbf{x}\in\mathbb{R}^{d+1}:\langle\mathbf{x}, \mathbf{x}\rangle_L=-K, x_0>0\}.
\end{equation}

Furthermore, we can use \textit{exponential mapping}, \textit{logarithmic mapping}, and \textit{parallel transport} equation for mapping between euclidean space and hyperbolic space:
\begin{equation}
    exp_\mathbf{x}^K(\mathbf{v})=cosh(\frac{\Vert\mathbf{v}\Vert _\mathcal{L}}{\sqrt{K}})\mathbf{x}+\sqrt{K} sinh({\frac{\Vert\mathbf{v}\Vert _\mathcal{L}}{\sqrt{K}}})\frac{\mathbf{v}}{\Vert\mathbf{v}\Vert_\mathcal{L}},
\end{equation}
where $\Vert\mathbf{v}\Vert_\mathcal{L}=\sqrt{\langle\mathbf{v}, \mathbf{v}\rangle_\mathcal{L}}$ denotes the norm of $\mathbf{v}$.
\begin{equation}
    log_\mathbf{x}^K(\mathbf{v})=d_\mathcal{L}^K(\mathbf{x}, \mathbf{v})\frac{\mathbf{v}+\frac{1}{K}\langle\mathbf{x},\mathbf{v}\rangle_\mathcal{L}\mathbf{x}}{\Vert\mathbf{v}+\frac{1}{K}\langle\mathbf{x},\mathbf{v}\rangle_\mathcal{L}\mathbf{x}\Vert_\mathcal{L}},
\end{equation}
where $d_\mathcal{L}^K(\mathbf{x}, \mathbf{v})$ is the distance between two points in $\mathbb{H}_d^K$:
\begin{equation}
    d_\mathcal{L}^K(\mathbf{x}, \mathbf{v})=\sqrt{K}arcosh(\frac{-\langle\mathbf{x},\mathbf{v}\rangle_\mathcal{L}}{K}).
\end{equation}

\textit{Parallel transport} in hyperboloid model is used for mapping a vector from one tangent space to another. Concretely, given a vector $\mathbf{v} \in \mathcal{T}_{\mathbf{x}\mathcal{M}}$, transport $P_{\mathbf{x}\to\mathbf{y}}(\mathbf{v})$ maps $\mathbf{v}$ to a Riemannian tensor in $\mathcal{T}_{\mathbf{y}}\mathcal{M}$.
\subsection{Problem Definition: Anomaly Detection on Attributed Networks}
\paragraph{Attributed Network.} Given a node set $\mathcal{V}=(v_1,...,v_n)$, an adjacency matrix $\mathbf{A}\in\mathbb{R}^{n\times n}$, and an attribute matrix $\mathbf{X}\in\mathbb{R}^{n\times d}$, an attributed network $\mathcal{G}$ can be defined as $\mathcal{G}=(\mathcal{V}, \mathbf{A}, \mathbf{X})$. The $i$-th row of attribute matrix $\mathbf{X}$ denotes the attribute information of node $v_i$. In the adjacency matrix, $\mathbf{A}_{i,j}=1$ if there is a connection between node $v_i$ and $v_j$, otherwise $\mathbf{A}_{i,j}=0$.
\paragraph{Anomaly Detection on Attributed Network.} Given an attributed network $\mathcal{G}=(\mathcal{V}, \mathbf{A}, \mathbf{X})$, our task is to learn a  function $f$ to calculate the anomaly score $f(v_i)$ for each node $v_i \in \mathcal{G}$, according to its attribute information and graph structure. The higher the node's anomaly score is, the more likely it is to be anomalous.

\subsection{Contrastive Learning for Anomaly Detection in Hyperbolic Space}\label{sec:method}
We propose an efficient hyperboloid contrastive learning method for anomaly detection. As shown in Figure \ref{fig:model}, our detection pipeline contains four steps: (1) We employ graph data augmentation by sampling positive and negative pairs for each node. (2) The hyperboloid encoder and decoder extract hierarchical embedding in the tangent space for the selected node and its neighbor embedding through a graph readout module. (3) We obtain discriminate score for positive and negative pairs through a discriminator respectively, measuring the agreement between the selected node with different pairs. (4) Finally, we calculate the discriminate scores mentioned above using multiple rounds sampling method, which is more robust to our paradigm. Specifically, it is an overall observation of all surrounding neighbors for each node.

In general, the training object is to discriminate positive pairs and negative pairs in the tangent space, which is highly related to the anomaly detection task, due to the disagreement between anomalous node and its neighbor.
\subsubsection{Graph Sampling}
The most important step of contrastive learning is data augmentation. Different from contrastive learning method applied in computer vision~\cite{chen2020simclr,he2020moco} and natural language processing~\cite{gao2021simcse}, graph data augmentation is more complex in specific method and implementation level. Existing graph data augmentation methods such as~\cite{zhu2021gca,mvgrl}, are mainly employed on nodes and edges at the whole graph level, which is useful for graph embedding learning but unsuitable for anomaly detection, especially in hyperbolic space, which may perturb the hierarchical order. Inspired by CoLA~\cite{liu2021cola}, an abnormal node usually shows more inconsistency with its neighbors in anomaly detection task. The heuristic is that a normal node is more consistent with its immediate neighbor than other non-neighbor nodes, but abnormal nodes never are. Thus, we extend a sampling method to hyperbolic space preserving its local hierarchical structure for anomaly detection.

Concretely, our designation focuses on the consistency of the node and its local neighbors in the tangent space via a ``target node v.s. neighbor subgraph" paradigm. As is shown in Figure \ref{fig:model}, first, a target node is randomly selected from the whole graph. Second, its neighbor is denoted as the node set by adopting RWR algorithm~\cite{tong2006rwr} in the tangent space. 
For a node $v_i$, a positive pair $P_i^+=\{v|v\in Neighbor(v_i)\}$ is the neighbor of $v$, and a negative pair $P_i^-=\{v|v\notin Neighbor(v_i)\}$ is randomly chosen from the neighbor of other nodes. To prevent contrastive learning method from recognizing such patterns easily, we also anonymize target node in the sampled pairs. Specifically, we mask the attribute vector of the target node.

Overall, we implement the graph traversal algorithm to sample for each point in the whole graph.
\subsubsection{Hyperboloid Transformation} After the graph sample step, instance pairs (either positive or negative) are denoted as $\mathcal{P}=(P^{+}_1,...,P^{+}_n, P^{-}_1,...,P^{-}_n)$, and $n$ is the number of nodes. For each instance pair $P_i=(v_i, \mathcal{S}_i, y_i)$, where $v_i$ is the selected target node, $\mathcal{S}_i=(A_i, X_i)$ is the subgraph generated from random walk with restart algorithm with node $v_i$, where $A_i$ denotes subgraph adjacency matrix, and $X_i$ denotes subgraph attribute matrix. $y_i$ denotes the node anomaly label, and is defined as: if $v_i \text{ is abnormal}$, then $y_i=1$; otherwise, $y_i=0$.

To utilize the rich hierarchical information in attributed graph, we first map the attribute of node $v_i$ and nodes in subgraph $\mathcal{S}_i=$ from the tangent space to hyperbolic space at the origin point by an \textit{exponential mapping}. For each node $v\in \{v_i \cup \mathcal{S}_i\}$, and its attribute $x$:
\begin{equation}
    \begin{split}
    x^{H}&=exp_{\mathbf{o}}^K((0, x^{E})),\\
    &=(\sqrt{K}cosh(\frac{\Vert x^{E}\Vert_2}{\sqrt{K}}) \sqrt{K}sinh(\frac{\Vert x^{E}\Vert_2}{\sqrt{K}})\frac{x^{E}}{\Vert x^{E}\Vert_2}),
    \end{split}
\end{equation}
where $(0,x^{E})\in\mathbb{R}^{d+1}$ is considered as a point in $\mathcal{T}_{\mathbf{o}}\mathbb{H}^{d,K}$, and $\mathbf{o}$ denotes the origin point in $\mathbb{H}^{d,K}$. Recall that $\langle(0,x^{E}), \mathbf{o}\rangle=0$, where we use $\mathbf{o}=(\sqrt{K},0,...,0) \in H^{d,K}$ as a reference point to perform tangent space operations.

For each $\mathcal{S}_i$, we apply graph convolution module to aggregate its local information in the subgraph and embed hierarchical information for each node simultaneously. A single layer $l$ with a learning curvature $K$ of hyperbolic hraph convolution can be written as:
\begin{align}
    H_i^l&=(W^{l-1}\otimes^{K_{l-1}}H_i^{l-1})\oplus^{K_{l-1}}b^{l-1}, \\
    AGG_i^l&=Aggregate^{K_{l-1}}(log_{\mathbf{o}}^{K_{l-1}}(H_i^{l-1}), Adj_i),\\
    Act_i^l&=\sigma^{K_{l, l-1}}(AGG_i^l),
    \label{equ:act}
\end{align}
where $H_i^{l-1}$ and $H_i^l$ $\in\mathbb{R}^{(subgraph\_size, d)}$ are the hidden subgraph representations of layer $l-1$ and layer $l$, $W_l$ and $b^l$ are the learning parameters, $Adj_i$ denotes the normalized adjacency matrix $\widetilde{D_i}^{-\frac{1}{2}}\widetilde{(A_i+I)}\widetilde{D_i}^{-\frac{1}{2}}$. For $l=0$, $H_i^0 = X_i^H$ is the subgraph attributed matrix in hyperbolic space. Operator $\otimes^{K_{l-1}}$ represents hyperbolic matrix multiplication, which can be computed through the \textit{logarithmic mapping} followed by the \textit{exponential mapping}:
\begin{equation}
    W^l\otimes^{K_{l-1}}H^{l-1}=exp_{\mathbf{o}}^{K_{l-1}}(Wlog_{\mathbf{o}}^{K_{l-1}}H^{l-1}).
\end{equation}
The same as $\otimes^{K_{l-1}}$, $\oplus^{K_{l-1}}$ is mobius add in hyperbolic space with curvature $K_{l-1}$, can be computed through the \textit{parallel transport}:
\begin{equation}
    H^{l-1}\oplus^{K_{l-1}}b^{l-1}=exp_{H^{l-1}}^{K_{l-1}}(P_{\mathbf{o}\to H^{l-1}}^{K_{l-1}}(b)).
\end{equation}
In detail, we implement the ReLU function as non-linear activation function in equation \eqref{equ:act}. 
\subsubsection{Anomaly Discriminator}
The discriminator aims to distinguish between positive and negative pairs for a selected node and its local subgraphs. Specifically, we first decode the node representation to the tangent sapce and then project it into classification space. Furthermore, we apply $bilinear$ function to compute the similarity score between a selected node and its subgraph instance pairs via a graph readout module.
\begin{equation}
    H_i^E=W\times log_{\mathbf{o}}^K(H_i^H)+b,
\end{equation}
\begin{equation}
    s_i=Readout(H_i^E)=\frac{1}{n}\sum_{k=0}^{n}h_k^E,
\end{equation}
where $n$ is the number of nodes in a subgraph. We can get the node representation for $v_i$ with the same hyperboloid encoder and decoder for the following score computation:
\begin{equation}
    q_i=Bilinear(DEC(ENC(X_{(v_i)})), s_i),
\end{equation}
where the \textit{ENC} denotes the convolutional encoder in our hyperboloid model mentioned above and the \textit{DEC} denotes the \textit{parallel transportation} from the hyperbolic space to the tangent space at the origin point.
It is worth mentioning that we also apply different kinds of discriminate methods, including contrasting positive and negative pairs in hyperbolic space directly and attention-based readout module. But we do not observe any improvement. Detailed results are discussed in the Ablation Study, where conducting \textit{w/o hyperboloid decoder} denotes discriminate directly in the hyperbolic space.
\subsubsection{Loss Function}
After discriminate score computation, our contrastive learning objective can be viewed as the binary classification task. The learning process is to discriminate between target node and positive or negative pairs. The same as label $y_i$, for a positive pair the predicted label is expected to be $1$, while for negative one is expected to be $0$. Hence, we adopt binary cross-entropy (BCE) loss as our criterion function. Given a batch of training instances $\mathcal{P}=\{\mathcal{P}_1,...,\mathcal{P}_N\}$, our criterion function can be computed as:
\begin{equation}
    \mathcal{L}=-\sum_{i=1}^N y_ilog(q_i) + (1 - y_i)log(1 - q_i).
    \label{eq:loss}
\end{equation}
\subsubsection{Anomaly Detection}
In the final step, we can get anomaly score by making subtraction between the score of a positive pair $\mathcal{P}_i^+$ and a negative pair $\mathcal{P}_i^-$, since a normal node is consistent with its neighbor in a positive pair, while exhibits more inconsistency with a negative pair. For a normal node, the discriminate score of a positive pair is closer to 1, while for a negative pair is closer to 0. In contrast, for a abnormal node, there is small margin of discriminate score between a positive and negative pair, because the model fails to generalize such pattern via training on a large amount of normal node. Hence, the anomaly score can be computed as:
\begin{equation}
    Q_i=q_i^- - q_i^+.
    \label{equ:score}
\end{equation}
Concretely, for further robustness of our hyperboloid model, we sample $R$ rounds for each node, obtaining an overall view. On the other hand, for a structural anomaly node which contains some uncorrelated edges in the subgraph is more difficult to capture its abnormality through few rounds detection. Thus, we implement the multi-round neighbor detection, which can be computed as:
\begin{align}
    Q_{(i,R)}&=q_{(i,r)}^- - q_{(i,R)}^+,\\
    Q_i&=\frac{1}{R}\sum_{k=0}^R Q_{(i,R)},
\end{align}
where $q_{(i,R)}$ denotes the discriminate score of node $v_i$ either a positive or negative instance pair.
\subsubsection{Time Complexity Analysis}
We further analyze the time complexity of our framework with main steps aforementioned in section \ref{sec:method}. 
Specifically, we apply random walk with restart (RWR) algorithm to sample positive and negative pairs, the time complexity for batch size $n$ with local subgraph size $c$ and mean degree $\sigma$ is $\mathcal{O}(nc\sigma)$. In the hyperboloid encoder, we perform hyperboloid linear transformation, local aggregation and non-linear activation, which costs $\mathcal{O}(nkc^2)$, where $k$ denotes the hidden size. In the hyperboloid decoder, we employ tangent space projection, which costs $\mathcal{O}(nc)$. Furthermore, the discriminator for each instance costs $\mathcal{O}(2nk)$. Finally, the overall time complexity is $\mathcal{O}(n(c\sigma+kc^2+c+2k))$.
\section{Experiments}

\textbf{Datesets}. To evaluate our proposed hyperboloid model, we test the performance on four widely used datasets, including three social networks and a flight network. The statistics of the datasets are listed in Table \ref{table:data}. We report Gromov's $\sigma$ hyperbolicity values to indicate its hyperbolic geometry, where the lower $\sigma$ is, the more hyperbolic the dataset should be.
\begin{itemize}
    \item \textbf{Citation Networks.} CORA, CITESEER and PUBMED are public citation network datasets. Nodes represent scientific papers and edges represent citation relations. The attribute feature is the bag of word representation, where dimension is determined by its dictionary size.
    \item \textbf{Flight Networks.} We follow the same settings in HGCN~\cite{chami2019hyperbolicgcn}. The nodes in airport denote the airports, and edges indicate flights. The attributes indicate the geographic information including longitude, latitude, altitude and GDP.
\end{itemize}

 Following previous works~\cite{ding2019dominant,liu2021cola}, we inject a combination of structural anomalies and attribute anomalies into each dataset since there are no ground-truth anomalies in these datasets mentioned above. Concretely, the attribute anomalies are constructed by adding extra noise to $k$ selected nodes. For node $v_i$, we first randomly select adequate number of nodes as candidates, and choose the least similar node $v_j$ to change the attribute of $v_i$ as $v_j$, according to the euclidean distance between two nodes. The structural anomalies are constructed by adding extra noise to the topological structure of the network. We first choose $n$ cliques that contains $m$ nodes each as fully connected graphs. To balance the number of two types of anomalies, we set $k=p\times q$, according to the number of nodes in each dataset. Specifically, following the settings in previous works, we set $p=15$ for all datasets, and $q=5,5,20,5$ for CORA, CITESEER, PUBMED, and AIRPORT.

\begin{table}[pt]
\small
\resizebox{\columnwidth}{!}{
\begin{tabular}{c|ccccc}
\hline
Dataset  & Nodes & Edges & Features & Anomaly & $\sigma$ \\ \hline
CORA     & 2708  & 5429  & 1433     & 0.06         & 2.5   \\
CITESEER & 3327  & 4732  & 3703     & 0.05         & 4.0   \\
PUBMED   & 19717 & 44338 & 500      & 0.03         & 2.5   \\
AIRPORT  & 3188  & 18631 & 1000     & 0.05         & 1.5   \\ \hline
\end{tabular}}
\vspace{-2mm}
\caption{Statistics of datasets.}
\vspace{-4mm}
\label{table:data}
\end{table}

\textbf{Baselines}.
For comparison, we choose different baseline models, including naive and localized (especially for anomaly detection task) contrastive learning models, and original hyperboloid models used for classification task. Since shallow based methods cannot achieve satisfactory performance limited by their intrinsic mechanism which cannot cope with complex network structure and sparse attributes, we do not make further comparison with these methods.
(1) \textbf{DOMINANT.}~\cite{ding2019dominant} DOMINANT is a reconstruction based method. It aims at reconstructing the adjacency matrix and the attribute matrix after employing graph convolution network. For inference step, it utilizes ranking strategy by choosing nodes with higher reconstruction error as anomalies.
(2) \textbf{MVGRL.}~\cite{mvgrl} MVGRL is a graph contrastive learning method for representation learning. It generates graph embedding by contrasting in ``node v.s node" and ``graph v.s graph" views simultaneously. The bilinear function is used to discriminate between positive and negative sample pairs.
(3) \textbf{CoLA.}~\cite{liu2021cola} CoLA is a contrastive learning method framework for anomaly detection task in euclidean space. It utilizes the general ``node v.s neighbor" paradigm but the hierarchical information is ignored in this process. For AIRPORT dataset, we take the same hidden dimension setting used in our model, and then take an average metric on different ten seeds. Concretely, we set $lr=0.001$ and $hidden\_dim=64$ and run $100$ epochs for training.
(4) \textbf{HGCN.}~\cite{chami2019hyperbolicgcn} HGCN is the Hyperboloid Graph Convolutional Network, which is designed for graph embedding tasks via supervised learning. Here, we employ the same train/val/test split method as ~\cite{chami2019hyperbolicgcn}. For AIRPORT, we set the train/val/test set proportion as $0.15/0.15/0.7$ respectively.

\textbf{Hyperparameter Settings}.
For all datasets, the learning rate is set as $0.03$. We apply Adam optimizer with $0.999$ momentum, and $0.0001$ weight decay value for CITESEER, PUBMED and AIRPORT, $0.0002$ for CORA. We should note that we also apply Riemannian Adam optimizer, but it fails to converge in this situation. Dropout is the key to prevent the model from overfitting for a certain dataset. We set dropout rate to 0.5 for CORA and 0.1 for CITESEER, PUBMED and AIRPORT. We should note that we set curvature $K=2.5$ for CORA, PUBMED and AIRPORT, $K=1.0$ for CITESEER. In the inference stage, we perform $256$ rounds detection method to get an overall view of a target node.

\textbf{Anomaly Detection Results and Analysis}.
We evaluate the performance of our model on four datasets compared with three baseline models in Table \ref{table:result}. Concretely, we employ the ROC-AUC metric, where ROC is the curve of true positive rate against false positive rate. AUC is the value of ROC curve under area, which demonstrates how confident our model is to make the right predictions. When the AUC value is closer to 1, it means the model has a better performance.
\begin{table}[t]
\resizebox{\columnwidth}{!}{
\begin{tabular}{c|c|cccc}
\toprule
\multicolumn{2}{c|}{Methods} & \textbf{CORA} & \textbf{CITESEER} & \textbf{PUBMED} & \textbf{AIRPORT} \\
\multicolumn{2}{c|}{Hyperbolicity} & $\mathbf{\sigma}$=2.5 & $\mathbf{\sigma}$=4.0     & $\mathbf{\sigma}$=2.5   & $\mathbf{\sigma}$=1.5    \\ \midrule
\multirow{3}{*}{Euclidean}
&DOMINANT & 0.8155& 0.8251& 0.8081& 0.6964        \\
&MVGRL & 0.6255& 0.4096& 0.6714& 0.4604        \\
&CoLA & 0.8827& \textbf{0.8968}& 0.9512& 0.7575        \\\midrule
\multirow{2}{*}{Hyperboloid}
&HGCN & 0.5285& 0.6984& 0.5966& 0.6221 \\ 
&Ours & \textbf{0.8914}& 0.7618& \textbf{0.9513}& \textbf{0.7805} \\
\bottomrule
\end{tabular}}
\vspace{-1mm}
\caption{AUC results of our proposed approach and baselines.}
\vspace{-2mm}
\label{table:result}
\end{table}
The average results are exhibited in Table \ref{table:result}, where we experiment on five seeds. Compared with the baselines, our model improves \textbf{5.42\%} using the AUC score on AIRPORT dataset, which has the lowest hyperbolicity value. Overall, we can get the following observations. Our model shows significant performance on AIRPORT dataset, which has a lower hyperbolicity value. Generally, the more hyperbolic the dataset is, the better performance it can obtain.

\begin{table}[h]
\resizebox{\columnwidth}{!}{
\begin{tabular}{cccccc}
\hline
\textbf{Ablation}         & AIRPORT & PUBMED & CORA &\textbf{Avg.}\\ \hline
w/o hyperboloid decoder & 0.7273 & 0.9386 & 0.8097 & 0.8252\\
w/o hyperbolic space   & 0.7590 & 0.9160 & 0.6464 & 0.7738         \\
Full model             & 0.7805 & 0.9513 & 0.8914 & 0.8744   \\ \hline
\end{tabular}}
\vspace{-1mm}
\caption{AUC results of ablation study on the different datasets.} 
\vspace{-1mm}
\label{table:ablation}
\end{table}

\textbf{The Ablation Study}. To illustrate the role of encoder and decoder of our hyperboloid contrastive learning model, we perform an ablation study on the AIRPORT, PUBMED and CORA datasets, where we outperform the baseline approaches. As shown in Table ~\ref{table:ablation}, compared to employing contrastive learning model directly in the hyperbolic space without a \textit{logarithmic mapping} decoder, the AUC score decreases by \textbf{4.92\%}. Furthermore, we explore the role of our hyperboloid model by discriminating positive and negative pairs in the euclidean space. The AUC score decreases by \textbf{10.06\%}.

\begin{figure}[h]
    \centering
    \includegraphics[scale=0.4]{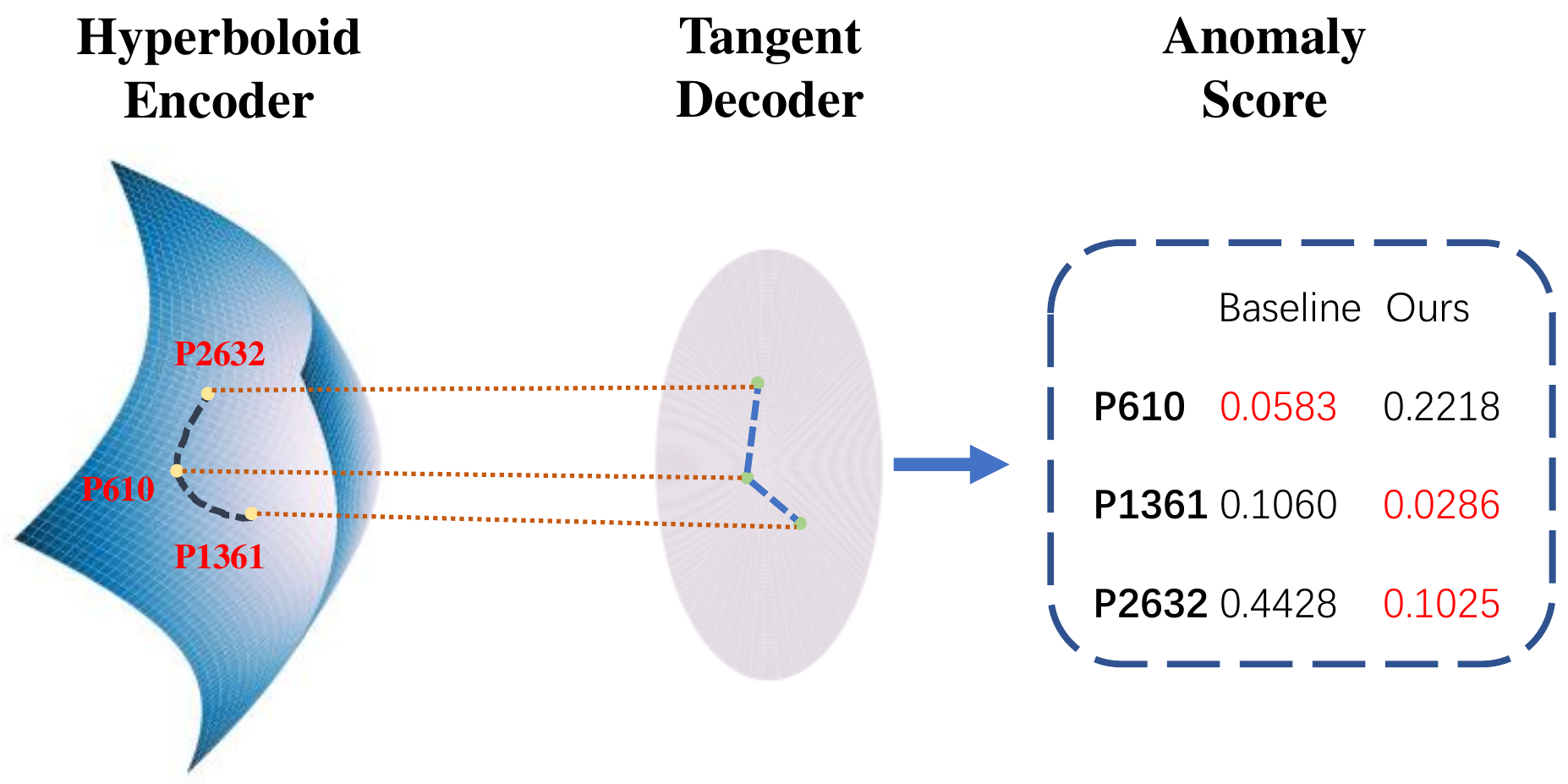}
    \vspace{-2mm}
    \caption{P610 and its anomaly neighbors in the AIRPORT dataset. We project the data point into the tangent space at the original point.}
    \vspace{-5mm}
    \label{fig:case}
\end{figure}

\textbf{Case Study}.
We conduct a case study on the AIRPORT dataset to make a detailed comparison with the Euclidean-based model CoLA~\cite{liu2021cola}. As shown in Figure~\ref{fig:case}, P610 has two \textbf{anomaly} neighbors P1361 and P2632(due to the space limitation, we only show two of its neighbors). In comparison to the baseline model, our hyperboloid model can more efficiently detect anomalies utilizing the hierarchical information. Table \ref{table:case} shows that our model predicts anomalies with a small margin computed by equation(\ref{equ:score}). The smaller the margin is, the more likely it is an anomaly node. Hence, we can imply that our model is much easier to distinguish anomalies from the normal ones than the baseline with {the same 32 hidden neurons} setting.
\vspace{-2mm}
\begin{table}[h]
\resizebox{\columnwidth}{!}{
\begin{tabular}{c|cccc}
\toprule
\textbf{Methods} & Point & Negative($q_i^-$) & Positive($q_i^+$) & Margin=$|q_i^- - q_i^+|$\\ \midrule
\multirow{3}{*}{CoLA}
&P610 & 0.2285& 0.2868 & \textbf{0.0583}      \\
&P1361 & 0.2739& 0.1679& 0.1060    \\
&P2632 & 0.2805& 0.7233& 0.4428      \\ \midrule
\multirow{3}{*}{Ours}
&P610 & 0.1901& 0.4119& 0.2218        \\
&P1361 & 0.2156& 0.2442& \textbf{0.0286}       \\
&P2632 & 0.1631& 0.2656& \textbf{0.1025}       \\
\bottomrule
\end{tabular}}
\vspace{-1mm}
\caption{Results of P160 and its neighbor in the AIRPORT dataset, where the margin denotes the anomaly score.}
\vspace{-1mm}
\label{table:case}
\end{table}

\vspace{-4mm}
\section{Conclusion}
In this paper, we propose the anomaly detection framework based on hyperbolic 
self-supervised contrastive learning. We are the first to introduce the contrastive learning method to the hyperbolic space, especially for the anomaly detection task. In our hyperboloid model, hierarchical information is included via exponential and logarithmic mapping. Results show that the more hyperbolic the dataset is, the more hierarchical information it can obtain. Furthermore, we conduct an ablation study to illustrate the key role of hyperboloid encoder and decoder. In comparison to  baselines, our approach detects anomalies in a small margin, implying that our model is more confident in the detection results. We hope that our work will inspire future work on hyperbolic graph learning.
\newpage

\bibliographystyle{named}
\bibliography{ijcai22}

\end{document}